\documentclass[10pt,final,doublecolumn]{IEEEtran}
\usepackage{amsmath}
\usepackage{latexsym}
\usepackage{graphicx}
\usepackage{bbding}
\usepackage{enumerate}
\usepackage{multicol}
\usepackage{color}
\IEEEoverridecommandlockouts
\begin{document}

\title{Statistical Precoder Design for Space-Time-Frequency Block
Codes in Multiuser MISO-MC-CDMA Systems}

\author{\authorblockN{
Xiaoming~Chen, \emph{Senior Member, IEEE}, and Xiumin~Wang,
\emph{Member, IEEE}
\thanks{Xiaoming~Chen (chenxiaoming@nuaa.edu.cn) is with the
College of Electronic and Information Engineering, Nanjing
University of Aeronautics and Astronautics, Nanjing, China.
Xiuming~Wang (wxiumin@hfut.edu.cn) is with the School of Computer
and Information, Hefei University of Technology, Hefei,
China.}}}\maketitle

\begin{abstract}
In this paper, we present a space-time-frequency joint block coding
(STFBC) scheme to exploit the essential space-time-frequency degrees
of freedom of multiuser MISO-MC-CDMA systems. Specifically, we use a
series of orthogonal random codes to spread the space time code over
several sub-carriers to obtain multi-diversity gains, while
multiuser parallel transmission is applied over the same
sub-carriers by making use of multiple orthogonal code channels.
Furthermore, to improve the system performance, we put forward to
linear precoding to the predetermined orthogonal STFBC, including
transmitting directions selection and power allocation over these
directions. We propose a precoder design method by making use of
channel statistical information in time domain based on the
Kronecker correlation model for the channels, so feedback amount can
be decreased largely in multi-carrier systems. In addition, we give
the performance analysis from the perspectives of diversity order
and coding gain, respectively. Moreover, through asymptotic
analysis, we derive some simple precoder design methods, while
guaranteeing a good performance. Finally, numerical results validate
our theoretical claims.
\end{abstract}

\begin{keywords}
Multiuser MISO-MC-CDMA, Space-Time-Frequency Block Code, Optimal
Precoder Design, Iterative Power Water-Filling, Channel Statistical
Information.
\end{keywords}

\IEEEpeerreviewmaketitle

\section{Introduction}
The initial works of Telatar \cite{MIMO1}, Foschini and Gans
\cite{MIMO2}, Tarokh \emph{et al}. \cite{MIMO3}, and Guey \emph{et
al}. \cite{MIMO4} have spurred considerable research on
multiple-input multiple-output (MIMO) systems. As proved by previous
literatures, MIMO systems can improve spectrum efficiency, enhance
link reliability, or a combination of both by deploying multiple
antennas at transmitter or receiver. Meanwhile, with the increase of
the demand on data service, orthogonal frequency division
multiplexing (OFDM), as a promising transmission technique, has been
adopted by several broadband communication standards, such as LTE-A
and 802.16m. Therefore, MIMO-OFDM is considered as a potential
choice for future wireless broadband communication systems
\cite{LTE-A} \cite{802.16m}.

For MIMO-OFDM for high-speed wireless communications, frequency
resource has not been fully exploited. Loosely speaking, the
independent fading property of sub-carriers has not been utilized
effectively. In addition, for cellular systems, MIMO-OFDM suffers
from the interference of the same sub-carrier from adjacent cells.
Fortunately, MIMO-MC-CDMA \cite{MIMO-MC-CDMA1}-\cite{MIMO-MC-CDMA3}
has the ability of obtaining frequency gain and mitigating intercell
interference simultaneously. Specifically, MC-CDMA spreads each
symbol to be transmitted, by an orthogonal random code, over several
sub-carriers according to the variable orthogonal spreading factor
(VOSF) before inverse fast Fourier transform (IFFT). At the
receiver, the same symbols over different sub-carriers are combined
by a certain mean to enhance the transmission reliability. Through
allocating an orthogonal identified code (OID) for each cell, the
interference from adjacent cells can also be avoided directly.
Hence, MIMO-MC-CDMA is another feasible choice for broadband
wireless communication systems.

Apart from frequency resource, spatial and temporal degrees of
freedom for MIMO-MC-CDMA can also be utilized to further improve
diversity gain by making use of special space time signaling
schemes, such as space time block coding. In \cite{STBC1}, the
space-time block coded MIMO-MC-CDMA system was studied, and a
Bayesian monte carlo multiuser receiver was proposed to improve the
overall performance. Moreover, the performance of the space-time
block coded MIMO-MC-CDMA system was analyzed, and the closed-form
symbol error probabilities over Nakagami-$m$ fading channels was
presented in \cite{STBC2}. Furthermore, channel-independent
precoding were introduced into MC-CDMA systems to improve the
performance \cite{BlindPrecoding1}-\cite{BlindPrecoding3}.

It is proved that linear precoding by exploiting the channel state
information (CSI) at the transmitter can significantly improve
system performance and reduce receive complexity. Specifically,
linear precoding mainly includes transmission direction selection in
spatial domain and power allocation under a given constraint based
on the available CSI \cite{FCSI1}-\cite{FCSI3}. In \cite{FullCSI},
the transmitter beamforming scheme combining adaptive modulation
based on the full CSI is proposed for the space-time block coded
OFDM. However, for a FDD system, it is difficult to convey the full
CSI from the receiver to the transmitter. Especially in
multi-carrier systems, the CSI feedback amount is unbearable. To
satisfy the constraint of the amount of feedback, linear precoding
based on a finite-resolution quantization codebook, such as
Grassmann codebook \cite{Granssmann1} \cite{Grassmann2} and vector
quantization (VQ) codebook \cite{VQ1} \cite{VQ2}, becomes a common
solution in single-carrier systems. In \cite{PrecodedOSTBC1}, the
design of Grassmann codebook in the MIMO system employing space-time
codes has been well studied. Furthermore, \cite{PreocodedOSTBC2}
extended Grassmann codebook design to the irregular MIMO system. In
\cite{PreocodedOSTBC3}, the performance of VQ codebook in the
multiuser multi-antenna system was investigated. Although linear
precoding based on quantization codebook can achieve a proper
tradeoff between system performance and feedback amount, it is
unsuitable to multi-carrier system, since the sum of feedback amount
over all sub-carriers is unbearable for a real system. In order to
reduce the feedback bits, in \cite{Interpolation}, an
interpolation-based precoding strategy for multi-carrier systems
with limited feedback was proposed, which obtains CSI of all
sub-carriers via linear weight sum of those of limited sub-carriers
according to the correlation of frequency domain. Instead of full
feedback, this method reduces the feedback amount at the cost of
system performance. In addition, a subspace trace method based on
Grassmann manifold theory \cite{Track} can be used to solve the
problem in multi-carrier systems, but it requires a magnitude of
computations to gain the varied direction of subspace. More
importantly, codebook based precoding is unfit for the future
broadband mobile communication under some conditions, because the
CSI may be outdated when the transmitter acquires it through
feedback link due to fast channel variation, resulting in
performance loss \cite{CSIUncertain}. Fortunately, channel
statistical information, e.g. channel mean or correlation, varies
slowly and can be used to design the optimal preocoder \cite{Mean}
\cite{Correlation}. Moreover, a precoder design method based on
channel angular domain for distributed antenna systems was proposed
in \cite{AngularDomain} by minimizing the pairwise error probability
of space time codes. However, these works mainly considered the
precoder design in single carrier MIMO systems, which is unsuitable
for multi-carrier systems. Specifically, precoder design in
single-carrier MIMO systems is based on the statistical information
in time domain. However, in multi-carrier MIMO systems, such as
MIMO-MC-CDMA systems, the precoder is designed using the frequency
domain information. Note that it is impractical to estimate the
statistical information over all sub-carriers. A feasible way is to
build the relation between the time domain and frequency domain
statistical information. To the best of the authors' knowledge, it
is still an open issue to pose the statistical information of all
sub-carrier from those of time domain multi-path channel in
multi-carrier systems.

In this paper, we consider a general multiuser MISO-MC-CDMA system.
We combine linear precoding and orthogonal STFBC to effectively
improve the performance of MISO-MC-CDMA systems. The major
contributions of this paper can be summarized as follows:
\begin{enumerate}

\item We build an analytical framework for channel statistical information
in time domain over each sub-carrier based on Kronecker correlation
model.

\item We obtain the optimal precoder design method based on the
statistical information by minimizing the PEP of orthogonal STFBC.
Meanwhile, we also present an iterative water-filling power
allocation algorithm, which can further optimize the performance
with respect to equal power allocation and single beam allocation.

\item We get simple precoder design methods without performance loss
in some special scenarios through asymptotic analysis as follows:

\begin{enumerate}

\item When the number of delay taps, the length of spread codes
and the transmit SINR are large, equal power allocation can achieve
the asymptotic optimal performance.

\item When the SINR is low, distributing all the power to the
spatial sub-channel with the largest gain is asymptotically optimal.

\end{enumerate}

\end{enumerate}

The rest of this paper is organized as follows. Section II provides
a general overview of the MISO-MC-CDMA system under consideration.
In Section III, we focus on the precoder design using channel
statistical information. For some special cases, we give the
corresponding precoder design methods with low complexity in Section
IV. Simulation results are discussed and analyzed in Section V and
we conclude the whole paper in Section VI.

\textit{Notation}: We use bold upper (lower) letters to denote
matrices (column vectors), $(\cdot)^*$ to denote conjugate
transpose, $(\cdot)^T$ to denote matrix transpose, $\textbf{I}_{k}$
to denote the $k\times k$ identity matrix, $E[\cdot]$ to denote
expectation, $\lambda_{i}\{\textbf{A}\}$ to denote the $i$th largest
singular values of $\textbf{A}$, $\|\cdot\|_{2}$ to denote the
matrix $l_2$-norm, $\|\cdot\|_{F}$ to denote the matrix Frobenius
norm, $\textmd{tr}(\cdot)$ to denote matrix trace, $\sim$ to denote
equality in distribution, $\otimes$ to denote Kronecker product
operator, and $\succeq$ to denote the matrix positive semi-definite
relation. The acronym i.i.d. means ``independent and identically
distributed" and pdf means ``probability density function".

\section{System Model}

\begin{figure}[h] \centering
\includegraphics [width=0.5\textwidth] {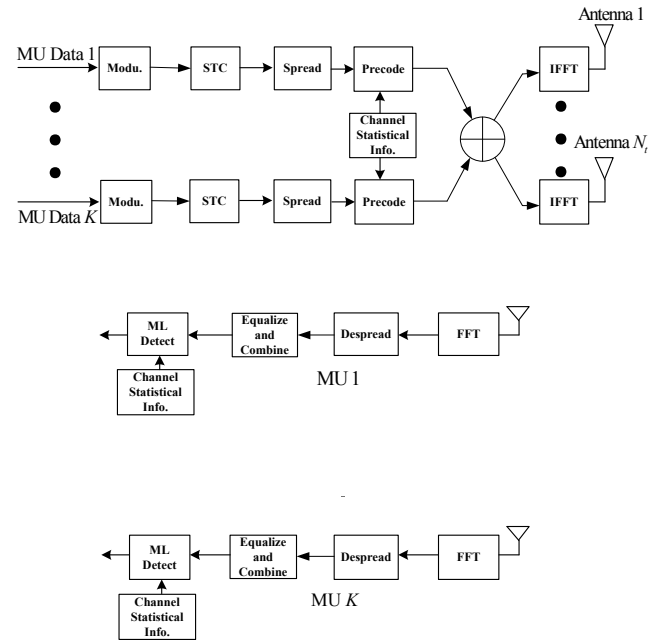}
\caption {A block diagram for MIMO-MC-CDMA systems with STC.}
\label{Fig1}
\end{figure}

Consider a multiuser MISO-MC-CDMA downlink, where the base station
(BS) has $N_t$ transmit antennas, and $K$ mobile users (MU) equip
with one receive antenna each, as illustrated in Fig.\ref{Fig1}.
MISO-MC-CDMA is a multiplexing technique which permits multiple
users to access the wireless channel simultaneously by modulating
and spreading their input data signals across the frequency domain
using different spreading sequences. Therefore, MISO-MC-CDMA
combines the advantages of MISO, OFDM and CDMA indeed. For the $k$th
MU, its data is first mapped into a series of modulated symbols by a
modulator. Then, the consecutive $M$ modulated symbols are used to
construct a space time code $\textbf{X}_{k}$ of size $M\times T$.
For analytical convenience, we consider orthogonal space block codes
in this paper. Next, the space time code is spread to $N_v$ parallel
codes by a VOSF
$\textbf{c}_{k}=\{c_{k,1},c_{k,2},\cdots,c_{k,N_v}\}$ of length
$N_v$. After the OFDM modulation with $N_c$ sub-carriers, the $N_v$
spread space time codes are mapped into $N_v$ different sub-carriers
with large spacing. Based on the statistical information of the
sub-carrier, the corresponding spread space time code is precoded by
the precoding matrix $\textbf{F}_{k,i}$ for $i=1,\cdots,N_v$. Due to
the VOSF of MIMO-MC-CDMA, multiple MUs' precoded space time codes
can be transmitted on a carrier simultaneously, so the spectrum
efficiency can be improved effectively. In this paper, for ease of
analysis, we assume each sub-carrier carries $N_u$ MUs' precoded
space time codes. Finally, the space time codes are transmitted by
using $T$ time slots. Assume the channel is block fading and
ergodic, in other words, the channel keeps constant during a space
time code of length $T$ time slots and fades independently between
codes. It is noticed that, due to the homogeneous channels of
different MUs, we normalize the path-dependent fading, including
path loss and shadow fading. Without loss of generality, we focus on
investigating the design of optimal precoding matrix of MU $0$ and
analyzing the corresponding performance throughout this paper. Thus,
the baseband frequency domain output signal of the $0$th MU on the
$i$th sub-carrier can be expressed as
\begin{equation}
\textbf{Y}_{0,i}=\sqrt{\rho}\textbf{H}_{0,i}\textbf{F}_{0,i}c_{0,i}\textbf{X}_{0}+\sqrt{\rho}\sum\limits_{k=1}^{N_u-1}\textbf{H}_{0,i}\textbf{F}_{k,i}c_{k,i}\textbf{X}_{k}+\textbf{N}_{0,i},\label{equ1}
\end{equation}
where $\rho$ is the total transmit power, $\textbf{F}_{0,i}$ is the
corresponding precoding matrix designed based on channel statistical
information. As mentioned above, $\textbf{F}_{0,i}$ can be expressed
as the product of a $N_t\times M$ precoding direction matrix
$\textbf{V}_{0,i}$ that is used to specify the waveform of space
time code and a $M\times M$ positive semi-definite diagonal power
allocation matrix $\textbf{D}_{0,i}$, $\textbf{X}_{0}$ and
$\textbf{Y}_{0,i}$ are $M\times T$ transmitted and $1\times T$
received space time codes (STC), respectively. $\textbf{N}_{0,i}$ is
$1\times T$ received noise vector with i.i.d entries
$\sim\mathcal{CN}(0,\sigma_{n}^2)$. $c_{0,i}$ is the spread code,
and $N_u$ is the number of users on one sub-carrier. Assume the VOSF
for all MU is $N_v$, viz. the same STC is spread over $N_v$
sub-carriers with orthogonal random code of length $N_v$. Thus, we
have
\begin{equation}
\sum\limits_{i=1}^{N_v}c_{k,i}c_{l,i}=\begin{cases}0, & k\neq l,\\
N_v, & k=l, \end{cases}\nonumber
\end{equation}
subject to that $N_c/N_v$ is an integer in order to effectively
utilize all the sub-carriers. $\textbf{H}_{0,i}$ is the $N_t$
dimensional channel response vector from the BS to MU 0 over the
$i$th sub-carrier. In this paper, we consider a frequency-selective
MIMO channel with $L$ resoluble delay taps in time domain, which can
be expressed as
\begin{equation}
\textbf{G}_0(t)=\sum\limits_{l=0}^{L-1}\textbf{G}_{0,l}\delta(t-\tau_l),\label{equ5}
\end{equation}
where $\textbf{G}_{0,l}$ is the $N_t$ dimensional channel response
vector of the $l$th path in time domain, which has the corresponding
propagation delay $\tau_l$. In particular, $\textbf{G}_{0,l}$ is a
zero mean complex Gaussian random vector with the correlation matrix
$\textbf{R}_0$, namely, $\textbf{G}_{0,l}$ can be expressed as
\begin{equation}
\textbf{G}_{0,l}=\textbf{G}_{\textrm{i.i.d}}\textbf{R}_0^{1/2},\quad
l=0,...,L-1,\nonumber
\end{equation}
where $\textbf{G}_{\textrm{i.i.d}}$ is a random vector
$\sim\mathcal{CN}\left(0,\textbf{I}_{N_t}\right)$. Thus, the
frequency response of the $n$th sub-carrier ($0\leq n\leq N_c$) can
be written as
\begin{eqnarray}
\textbf{H}_{0,n}&=&\sum\limits_{l=0}^{L-1}\textbf{G}_{0,l}\exp\left(-j2\pi
nl/N_c\right)\nonumber\\
&=&\textbf{G}_{0,\textrm{path}}\cdot\left(\textbf{W}_n\otimes\textbf{I}_{N_t}\right),\label{equ6}
\end{eqnarray}
where
$\textbf{G}_{0,\textrm{path}}=\left[\textbf{G}_{0,0},...,\textbf{G}_{0,L-1}\right]$
and $\textbf{W}_n=\left[\exp\left(-j2\pi n\times
0/N_c\right),...,\exp\left(-j2\pi n\times
(L-1)/N_c\right)\right]^T$. As seen in (\ref{equ6}),
$\textbf{H}_{0,n}$ is a sum of $\textbf{G}_{0,l}$ with linear
weight, so that $\textbf{H}_{0,n}$ is also a zero mean complex
Gaussian random vector, whose correlation matrix can be casted as
\begin{eqnarray}
\textbf{R}_{0,n}&=&E\left[\textbf{H}_{0,n}^{H}\textbf{H}_{0,n}\right]\nonumber\\
&=&\left(\textbf{W}_n\otimes\textbf{I}_{N_t}\right)^{H}E\left[\textbf{G}^{H}_{0,\textrm{path}}\textbf{G}_{0,\textrm{path}}\right]\left(\textbf{W}_n\otimes\textbf{I}_{N_t}\right)\label{equ7}\\
&=&\left(\textbf{W}_n\otimes\textbf{I}_{N_t}\right)^{H}\left(\textbf{R}_{0,\textrm{tap}}^{T}\otimes\textbf{R}_0\right)\left(\textbf{W}_n\otimes\textbf{I}_{N_t}\right)\label{equ8}\\
&=&\left(\textbf{W}_n^{H}\textbf{R}_{0,\textrm{tap}}^{T}\textbf{W}_n\right)\otimes\textbf{R}_0\label{equ9}\\
&=&\textbf{W}_n^{H}\textbf{R}_{0,\textrm{tap}}^{T}\textbf{W}_n\textbf{R}_0,\label{equ10}
\end{eqnarray}
where $\textbf{R}_{0,\textrm{tap}}$ is the tap correlation matrix.
In this paper, we assume that tap correlation is independent of
transmit correlation. Equation (\ref{equ9}) uses the Kronecker
product property that
$(\textbf{A}\otimes\textbf{B})(\textbf{C}\otimes
\textbf{D})=(\textbf{AC})\otimes(\textbf{BD})$. For analytical
simplicity, we assume that the taps are independent of each other,
namely, $\textbf{R}_{0,\textrm{tap}}=\textbf{I}_L$. In fact, if each
path has its unique propagation environment, it is reasonable to
assume the taps are uncorrelated. Under this condition, we get an
interesting result that the correlation matrices of all sub-carrier
are identical for an arbitrary MU. Therefore, the distribution of
the $n$th sub-carrier can be written as
\begin{equation}
\textbf{H}_{0,n}\sim\mathcal{CN}\left(0,L\textbf{R}_0\right).\label{equ11}
\end{equation}

Based on channel statistical information in time domain, such as
channel mean, channel covariance, and tap correlation matrices,
which can be obtained by averaging a number of channel realization,
the statistical information of all sub-carriers can be figured out
easily. Due to its slow variety, channel statistical information
keeps constant during a relatively long time, which promises its
advantage over instantaneous channel information for future
broadband wireless communication systems with limited feedback. In
this paper, we assume perfect channel statistical information at the
BS and MUs.

As seen in (\ref{equ1}), the first term in R.H.S is the expected
received signal, the middle term is the other $N_u-1$ MUs'
interference on the same sub-carrier, and the third term is the
received noise. To enhance the transmission reliability, namely
increasing received SINR, it is imperative to suppress the
multi-user interference (MUI). Inspired by the orthogonality of
spread code between MUs on the same set of sub-carriers, we first
multiply the received signal by the corresponding MU's spread code
to filter MUI, then combine the same transmitted signals on $N_v$
different sub-carriers. Thus, the combined signal of the $0$th MU
can be written as
\begin{eqnarray}
\bar{\textbf{Y}}_{0}&=&\sqrt{\rho}\textbf{X}_{0}\sum\limits_{i=0}^{N_v-1}
\textbf{H}_{0,i}\textbf{F}_{0,i}\nonumber\\&&+\sqrt{\rho}\sum\limits_{k=1}^{N_u-1}
\textbf{X}_{k}\sum\limits_{i=0}^{N_v-1}\textbf{H}_{0,i}\textbf{F}_{k,i}c_{k,i}c_{0,i}+\bar{\textbf{N}}_{0},\label{equ2}
\end{eqnarray}
where
$\bar{\textbf{N}}_{0}=\sum\limits_{i=0}^{N_v-1}c_{0,i}\textbf{N}_{0,i}$
is the combined noise with i.i.d. entries
$\sim\mathcal{CN}(0,N_v\sigma_n^2)$. Because of frequency selective
fading, the orthogonality of code channel is destroyed, which
results in that MUI can not be canceled completely. One feasible
solution to combat residual MUI is to adopt equalization technique
before combination, in (\ref{equ2}), we use equal gain combination
(EGC) instead of other complicated methods due to its simplicity.

After EGC in frequency domain, maximal-likelihood (ML) detection is
performed to obtain
\begin{equation}
\hat{\textbf{X}}_{0} = \arg \mathop {\min }\limits_{\textbf{X}_{0}
\in {\cal X}} \left\| {\bar{\textbf{Y}}_{0} - \sqrt \rho
{\textbf{X}_{0}\sum\limits_{i=0}^{N_v-1}\textbf{H}_{0,i}\textbf{F}_{0,i}}}
\right\|_F,\label{equ3}
\end{equation}
where $\cal{X}$ is the STC codebook. According to the law of large
number (LLN), if the term $N_vN_u$ is large enough, the second term
in (\ref{equ2}), namely MUI, can be considered as a zero mean
Gaussian random variable with the covariance of $\sigma_{MUI}^2$.
Intuitively, $\sigma_{MUI}^2$ is a function of $N_u$, $N_v$ and
$\rho$. $\sigma_{MUI}^2$ can be computed by statistical average on a
large number of samples, for given $N_u$, $N_v$ and $\rho$. Thus,
for a given channel realization
$\textbf{H}_0=[\textbf{H}_{0,1},\cdots,\textbf{H}_{0,N_v}]$, with ML
detection in (\ref{equ3}) and applying the Chernoff bound, the PEP
of precoded STFBC in MIMO-MC-CDMA systems can be tightly upper-bound
\cite{BeOSTBC} by (\ref{equ4}) at the top of next page.
\begin{figure*}
\begin{equation}
P\left( {{\bf X}_0 \to {\bf \hat X}_0}|\textbf{H}_0 \right) \le \exp
\left( { - \frac{\rho{\left\|
{\sum\limits_{i=0}^{N_v-1}\textbf{H}_{0,i}\textbf{F}_{0,i}\left(
{{\bf X}_0 - {\bf \hat X}_0} \right)} \right\|_F^2
}}{4\left(N_v\sigma_{n}^2+\sigma_{MUI}^2\right)}}
\right).\label{equ4}
\end{equation}
\end{figure*}

The focus of this paper is on the design of the optimal precoder
$\textbf{F}_{0,i},i=0,...,N_v$ at the BS based on channel
statistical information. The precoder can be divided into two
components: the $N_t \times M$ unitary matrix $\textbf{V}_{0,i}$ and
the $M \times M$ diagonal positive semi-definite matrix
$\textbf{D}_{0,i}$. The matrix $\textbf{V}_{0,i}$, as the precoding
direction information, serves to adapt the predetermined space time
codeword to the current channel condition. Given the optimization
goal, e.g. minimizing the upper bound of PEP in this paper, the
matrix $\textbf{V}_{0,i}$ is a function of channel statistical
information on the $i$th sub-carrier. As another important precoding
information, the matrix $\textbf{D}_{0,i}$ is used to allocate
transmit power for the basis-beams of the precoding direction matrix
to further improve system performance.

\section{Optimal Precoding for 3D Block Code}
In this section, we investigate the precoder design for 3D block
codes (STFBC) in a multiuser MISO-MC-CDMA system based on channel
statistical information in detail. In short, by minimizing the tight
upper bound of average PEP of precoded 3D block code, we design the
optimal precoding direction matrix and present some effective power
allocation algorithms.

\subsection{Optimization Objective}
Prior to discussing the design of the optimal precoder, we first put
the attention on the performance criterion. As mentioned above, we
take the average PEP of precoded 3D block code as the optimization
objective. As seen in (\ref{equ4}), given the channel realization,
the upper bound can be expressed as (\ref{equ12}) at the top of next
page,
\begin{figure*}
\begin{eqnarray}
P\left(\sum\limits_{i=0}^{N_v-1}\textbf{H}_{0,i}\right)&=&\exp
\left( { - \frac{\rho{\left\|
{\sum\limits_{i=0}^{N_v-1}\textbf{H}_{0,i}\textbf{F}_{0,i}\left(
{{\bf X}_0 - {\bf \hat X}_0} \right)} \right\|_F^2
}}{4\left(N_v\sigma_{n}^2+\sigma_{MUI}^2\right)}}
\right)\nonumber\\
&=&\exp
\left(-\eta\textrm{tr}\left(\left(\sum\limits_{i=0}^{N_v-1}\textbf{H}_{0,i}\right)
\textbf{F}_{0}\textbf{A}_0\textbf{F}_{0}^{H}\left(\sum\limits_{i=0}^{N_v-1}\textbf{H}_{0,i}\right)^{H}\right)\right).\label{equ12}
\end{eqnarray}
\end{figure*}
where
$\eta=\frac{\rho}{4\left(N_v\sigma_{n}^2+\sigma_{MUI}^2\right)}$,
and $\textbf{A}_0=\left( {{\bf X}_0 - {\bf \hat X}_0} \right)\left(
{{\bf X}_0 - {\bf \hat X}_0} \right)^{H}$ is the codeword distance
product matrix. (\ref{equ12}) results from the fact that the
precoder on an arbitrary sub-carrier is just a function of channel
statistical information and the space time codeword to the
transmitted, thus, based on the obtained conclusion above, the
precoders on all sub-carriers $\textbf{F}_{0,i}$ are identical for a
MU, namely $\textbf{F}_0$. As seen in (\ref{equ11}), the pdf of
sub-carrier $i$ can be expressed as
\begin{equation}
f\left(\textbf{H}_{0,i}\right)=\frac{\exp\left(-\textrm{tr}\left(\textbf{H}_{0,i}\textbf{M}_{0}^{-1}\textbf{H}_{0,i}^{H}\right)\right)}
{\pi^{N_t}\det\left(\textbf{M}_{0}\right)}, i=0,...,N_v-1,
\label{equ13}
\end{equation}
where $\textbf{M}_{0}=L\textbf{R}_0$ is the common covariance matrix
of MU $0$ on all $N_v$ chosen sub-carriers and $R_0$ is the
correlation matrix in time domain. Since the frequency interval
between adjacent sub-carriers is greater than correlation bandwidth,
it is reasonable to assume that the distributions of these
sub-carriers are independent of each other. Therefore, the pdf of
$\Phi=\sum\limits_{i=0}^{N_v-1}\textbf{H}_{0,i}$ is also a complex
Gaussian pdf, given by
\begin{equation}
f_{\Phi}\left(\textbf{Z}_{0}\right)=\frac{\exp\left(-\textrm{tr}\left(\textbf{Z}_{0}\textbf{P}_{0}^{-1}\textbf{Z}_{0}^{H}\right)\right)}{\pi^{N_t}\det\left(\textbf{P}_{0}\right)},\label{equ14}
\end{equation}
where $\textbf{P}_0=LN_v\textbf{R}_0$. Averaging (\ref{equ12}) over
(\ref{equ14}), we obtain the following bound on the average PEP:
\begin{eqnarray}
\bar{P}&=&\int_{\Phi}P\left(\textbf{Z}_{0}\right)f_{\Phi}\left(\textbf{Z}_{0}\right)
d\textbf{Z}_{0}\nonumber\\
&=&\frac{\int_{\Phi}\exp\left(-\textrm{tr}\left(\textbf{Z}_{0}\left(\eta\textbf{F}_{0}\textbf{A}_0\textbf{F}_{0}^{H}+\textbf{P}_{0}^{-1}\right)\textbf{Z}_{0}^{H}\right)\right)d\textbf{Z}_{0}}{\pi^{N_t}\det\left(\textbf{P}_{0}\right)}\nonumber\\
&=&\frac{\det\left(\textbf{Q}_{0}\right)}{\det\left(\textbf{P}_{0}\right)}\int_{\Phi}\frac{\exp\left(-\textrm{tr}\left(\textbf{Z}_{0}\textbf{Q}_{0}^{-1}\textbf{Z}_{0}^{H}\right)\right)}{\pi^{N_t}\det\left(\textbf{Q}_{0}\right)}d\textbf{Z}_{0}\label{equ16}\\
&=&\det\left(\textbf{I}_{N_t}+\eta
LN_v\textbf{F}_{0}\textbf{A}_0\textbf{F}_{0}^{H}\textbf{R}_{0}\right)^{-1},\label{equ17}
\end{eqnarray}
where
$\textbf{Q}_{0}=\left(\eta\textbf{F}_{0}\textbf{A}_0\textbf{F}_{0}^{H}+\textbf{P}_{0}^{-1}\right)^{-1}$,
(\ref{equ17}) is a result of that
$\int_{\Phi}\frac{\exp\left(-\textbf{Z}_{0}\textbf{Q}_{0}^{-1}\textbf{Z}_{0}^{H}\right)}{\pi^{N_t}\det\left(\textbf{Q}_{0}\right)}d\textbf{Z}_{0}$
is the integral of a complex Gaussian pdf and thus equals to one.
Now, we have got the exact expression of the upper bound of the
average PEP. In next sections, we start to investigate the design of
optimal precoder in the sense of statistical average.

\subsection{Optimal Precoder Design}
With the obtained the upper bound of the average PEP in
(\ref{equ17}), we construct the optimal precoder $\textbf{F}_{0}$,
including precoding direction matrix $\textbf{V}_0$ and power
allocation matrix $\textbf{D}_0$. As seen in (\ref{equ17}), the
optimal precoder is a function of channel correlation matrix
$\textbf{R}_0$ and space time codeword distance matrix $\textbf{A}$.
Thus, the design objective can be expressed as
\begin{equation}
\textbf{F}_0=\arg\max\limits_{\textbf{F}_0\in\textbf{C}^{N_t\times
M}}\det\left(\textbf{I}_{N_t}+\eta
LN_v\textbf{F}_{0}\textbf{A}_0\textbf{F}_{0}^{H}\textbf{R}_{0}\right),\label{equ18}
\end{equation}
where $\textbf{C}^{N_t\times M}$ denotes the set of complex matrices
of size $N_t\times M$.

As defined above, the precoder $\textbf{F}_0$ can be written as the
following form via singular value decomposition (SVD)
\begin{equation}
\textbf{F}_0=\textbf{V}_0\textbf{D}_0\textbf{U}_0^H,\label{equ19}
\end{equation}
where $\textbf{U}_0$ is an $M\times M$ unitary matrix. Similarly, we
give the SVD terms of $\textbf{A}_0$ and $\textbf{R}_0$ as
\begin{equation}
\textbf{A}_0=\textbf{V}_{\textbf{A},0}\textbf{D}_{\textbf{A},0}\textbf{V}_{\textbf{A},0}^H,\label{equ20}
\end{equation}
\begin{equation}
\textbf{R}_0=\textbf{V}_{\textbf{R},0}\textbf{D}_{\textbf{R},0}\textbf{V}_{\textbf{R},0}^H,\label{equ21}
\end{equation}
where $\textbf{V}_{\textbf{A},0}$ and $\textbf{V}_{\textbf{R},0}$
are $M\times M$ and $N_t\times N_t$ unitary matrices, respectively.
$\textbf{D}_{\textbf{A},0}$ and $\textbf{D}_{\textbf{R},0}$ are the
corresponding $M\times M$ and $N_t\times N_t$ positive semi-definite
diagonal matrices, respectively. It is well known that orthogonal
space time block code (OSTBC) is appealing in multi-antenna system
due to its lower decoding complexity. In what follows, we discuss
the design of optimal precoder based on such a code.

For OSTBC, it has the appealing property that
\begin{equation}
\textbf{A}_0=\mu_0\textbf{I}_{N_t},\label{equ22}
\end{equation}
where $\mu_0$ is a distance factor which depends on the modulation
mode of the symbol to be transmitted. Substituting (\ref{equ19}),
(\ref{equ21}) and (\ref{equ22}) into (\ref{equ18}) yields the
corresponding optimization objective for OSTBC
\begin{eqnarray}
(\textbf{V}_0, \textbf{D}_0,
\textbf{U}_0)&=&\arg\max\limits_{\textbf{V}_0\in\textbf{U}^{N_t\times
N_t},\textbf{D}_0\in\textbf{D}^{M\times M},
\textbf{U}_0\in\textbf{U}^{M\times
M}}\det(\textbf{I}_{N_t}\nonumber\\&&+\eta\mu_0
LN_v\textbf{V}_{0}\textbf{D}_{0}\textbf{U}_{0}^{*}\textbf{U}_{0}\textbf{D}_{0}\textbf{V}_{0}^{H}\textbf{V}_{\textbf{R},0}\textbf{D}_{\textbf{R},0}\textbf{V}_{\textbf{R},0}^H)\nonumber\\
&=&\arg\max\limits_{\textbf{V}_0\in\textbf{U}^{N_t\times
N_t},\textbf{D}_0\in\textbf{D}^{M\times M},
\textbf{U}_0\in\textbf{U}^{M\times
M}}\det(\textbf{I}_{N_t}\nonumber\\&&+\eta\mu_0
LN_v\textbf{D}_{0}^{2}\textbf{V}_{0}^{H}\textbf{V}_{\textbf{R},0}\textbf{D}_{\textbf{R},0}\textbf{V}_{\textbf{R},0}^H\textbf{V}_{0}),\label{equ23}
\end{eqnarray}
where $\textbf{U}^{N_t\times N_t}$ denotes the set of unitary
matrices of size $N_t\times N_t$ and $\textbf{D}^{M\times M}$
denotes the set of positive semi-definite diagonal matrices of size
$M\times M$. According to Hadamard's inequality \cite{Matrix}, for a
positive semi-definite matrix $\textbf{Q}$, its determinant
satisfies $\det(\textbf{Q})\leq\prod_i\|\textbf{q}_i\|_2$, where
$\textbf{q}_i$ is the $i$th column of $\textbf{Q}$. The equality
holds if and only if $\textbf{Q}$ is a diagonal matrix. Therefore,
in order to maximize (\ref{equ23}),
$\textbf{D}_{0}^{2}\textbf{V}_{0}^{H}\textbf{V}_{\textbf{R},0}\textbf{D}_{\textbf{R},0}\textbf{V}_{\textbf{R},0}^H\textbf{V}_{0}$
must be a diagonal matrix, namely
$\textbf{V}_0=\textbf{V}_{\textbf{R},0}$. As seen in (\ref{equ23}),
another optimization matrix $\textbf{U}_0$ is independent of the
objective function, so it can be an arbitrary unitary matrix, such
as $\textbf{I}_M$. Thus, the optimization objective reduces to
\begin{equation}
\textbf{D}_0=\arg\max\limits_{\textbf{D}_0\in\textbf{D}^{M\times
M}}\det\left(\textbf{I}_{N_t}+\eta\mu_0
LN_v\textbf{D}_{0}^{2}\textbf{D}_{\textbf{R},0}^{'}\right),\label{equ24}
\end{equation}
where $\textbf{D}_{\textbf{R},0}^{'}$ is a diagonal matrix composed
of the first $M$ diagonal elements of $\textbf{D}_{\textbf{R},0}$.
Let $\textbf{D}_0=\textmd{diag}\{d_1,\cdots,d_i,\cdots,d_M\}$ and
$\textbf{D}_{\textbf{R},0}^{'}=\textmd{diag}\{d_{\textbf{R},1},\cdots,d_{\textbf{R},i},\cdots,d_{\textbf{R},M}\}$,
where $d_i\geq0$ and $d_{\textbf{R},i}\geq0$ for $1\leq i\leq M$,
then the power allocation matrix $\textbf{D}_{0}$ can be obtained by
solving the following optimization problem
\begin{eqnarray}
J_1: \min\limits_{\textbf{D}_0}
&&-\prod\limits_{i=1}^{M}\bigg(1+\eta\mu_0
LN_vd_{\textbf{R},i}d_i^2\bigg)\nonumber\\
s.t.\quad &&\sum\limits_{i=1}^Md_i^2=1\nonumber\\
&&d_i\geq0, 1\leq i\leq M.\nonumber
\end{eqnarray}
Clearly, both the objective function and constraint conditions are
convex functions of optimization variables $d_i$, so $J_1$ is a
standard convex optimization problem. Form the Lagrangian of $J_1$
as
\begin{eqnarray}
L(\nu,\lambda_1,\cdots,\lambda_M)&=&-\prod\limits_{i=1}^{M}\bigg(1+\eta\mu_0
LN_vd_{\textbf{R},i}d_i^2\bigg)\nonumber\\&+&\nu\bigg(\sum\limits_{i=1}^Md_i^2-1\bigg)-\sum\limits_{i=1}
^{M}\lambda_id_i,
\end{eqnarray}
where $\nu$ is the Lagrange multiplier associated with the equality
constraint $\sum\limits_{i=1}^Md_i^2=1$, and $\lambda_i$ is the
Lagrange multiplier associated with the inequality constraint
$d_i\geq0$. Since $J_1$ is a convex optimization problem, strong
duality holds. Therefore, the primal and dual optimal points of
$J_1$ satisfy the following Karush-Kuhn-Tucher (KKT) conditions
\begin{eqnarray}
\sum\limits_{i=1}^M(d_i^*)^2&=&1\nonumber\\
d_i^*&\geq&0\nonumber\\
\lambda_i&\geq&0\nonumber\\
\lambda_i^*d_i^*&=&0\nonumber\\
\frac{\partial L(\nu,\lambda_1,\cdots,\lambda_M)}{\partial
d_i}\bigg|_{d_i^*,\nu^*,\lambda_i^*}&=&0,\nonumber
\end{eqnarray}
where $i=1,\cdots,M$, and $(\cdot)^*$ denotes the optimal value. By
solving the above KKT conditions, we have
\begin{equation}
d_i^*=\sqrt{\max\left(\frac{\eta\mu_0
LN_vd_{\textbf{R},i}-\nu^*}{\eta\mu_0
LN_vd_{\textbf{R},i}\nu^*},0\right)},
\end{equation}
where $\nu^*$ can be derived by solving the equality
$\sum\limits_{i=1}^M(d_i^*)^2=1$ through numerical iterative method.
Intuitively, the above power allocation follows the water-filling
principle \cite{Water-Filling}, we name it water-filling power
allocation in the rest of this paper.

%
%
%
%
%
%
%

\subsection{Performance Analysis}
Diversity order and coding gain are two most important indicatives
of SNR performance of various transmission schemes. In this section,
we derive the diversity order and coding gain of the precoded
space-time-frequency block code in multiuser MISO-MC-CDMA systems.
Diversity order $G_d$ and coding advantage $G_c$ are respectively
defined as the slope and offset of average PEP curves when SNR
$\eta$ approaches infinity as expressed below
\begin{equation}
G_d =  - \mathop {\lim }\limits_{\eta  \to \infty } \frac{{\log
\left( {\bar{P}\left( \eta  \right)} \right)}}{{\log \left( \eta
\right)}},
\end{equation}
and
\begin{equation}
G_c=\lim\limits_{\eta \to \infty}
\frac{\bar{P}^{-1/G_d}(\eta)}{\log(\eta)}.
\end{equation}

\textit{Theorem 1}: Precoded space-time-frequency block code in
multiuser MISO-MC-CDMA systems based on channel statistical
information can achieve full diversity order $N_t$.

\begin{proof} Substituting the optimal precoder into the upper bound
of the average PEP in (\ref{equ17}), we have
\begin{eqnarray}
\bar{P}&=&\det\left(\textbf{I}_{N_t}+\eta
LN_v\textbf{D}_{0}^{2}\textbf{D}_{\textbf{A},0}\textbf{D}_{\textbf{R},0}^{'}\right)^{-1}\nonumber\\
&\leq&\det\left(\eta
LN_v\textbf{D}_{0}^{2}\textbf{D}_{\textbf{A},0}\textbf{D}_{\textbf{R},0}^{'}\right)^{-1}\label{equ27}\\
&=&\eta^{-N_t}\det\left(LN_v\textbf{D}_{0}^{2}\textbf{D}_{\textbf{A},0}\textbf{D}_{\textbf{R},0}^{'}\right)^{-1},\label{equ28}
\end{eqnarray}
where (\ref{equ27}) results from the fact that
$\textbf{D}_{0}^{2}\textbf{D}_{\textbf{A},0}\textbf{D}_{\textbf{R},0}^{'}$
is a positive semi-definite matrix. Hence the upper bound of the
average PEP decays as $\eta^{-N_t}$, which proves the claim about
full diversity order in theorem 1. In this paper, we consider the
case with single receive antenna, however, our proposed transmission
scheme can be extended to the scenario with $N_r$ receive antennas
easily. Then, the diversity order is $N_tN_r$ provably. In addition,
multiuser diversity seems to be another available resource that
could be exploit through MU selection to increase diversity gains
\cite{Multiuser}.
\end{proof}

\textit{Theorem 2}: Precoded space-time-frequency block code in
multiuser MISO-MC-CDMA system based on channel statistical
information can pose $LN_v$ times coding gain than conventional
precoding space-time block code.

\begin{proof}
According to the definition of coding gain, the corresponding coding
gain of precoded space-time-frequency block code in multiuser
MIMO-MC-CDMA system based on channel statistical information can be
expressed as \cite{GD}
\begin{eqnarray}
G_c&=&\lim\limits_{\eta \to \infty}
\frac{\bar{P}^{-1/G_d}(\eta)}{\log(\eta)}\nonumber\\
&=&LN_v\det\left(\textbf{D}_{0}^{2}\textbf{D}_{\textbf{A},0}\textbf{D}_{\textbf{R},0}^{'}\right)^{1/G_d}\label{equ33}\\
&=&LN_v\left(\Pi_{m=1}^{M}d_{m}^{2}d_{\textbf{A},m}d_{\textbf{R},m}^{'}\right)^{1/G_d}.\label{equ34}
\end{eqnarray}
Built on the above analysis, it is clearly that, for single carrier
system, the coding gain is
$\det\left(\textbf{D}_{0}^{2}\textbf{D}_{\textbf{A},0}\textbf{D}_{\textbf{R},0}^{'}\right)^{1/G_d}$.
Thereby, through space-time-frequency coding, the coding gain is
improved $LN_v$ times.
\end{proof}

In addition, from (\ref{equ34}), it is found that, when $M=N_t$,
lower transmit correlation denotes higher coding gain. This is
because that, given
$\textmd{tr}(\textbf{R}_0)=\sum_{m=1}^{N_t}d_{\textbf{R},m}^{'}=N_t$,
$\Pi_{m=1}^{N_t}d_{\textbf{R},m}^{'}\leq\left(\frac{\sum_{m=1}^{N_t}d_{\textbf{R},m}^{'}}{N_t}\right)^{N_t}=1$
and the equality holds if and only if
$d_{\textbf{R},1}^{'}=\cdots=d_{\textbf{R},N_t}^{'}=1$. In other
words, the closer the eigenvalues are, the larger the coding gain
is, which means lower correlation is preferred. On the other hand,
for $M<N_t$, the best condition is
$d_{\textbf{R},1}^{'}=\cdots=d_{\textbf{R},M}^{'}=\frac{N_t}{M}$.
Therefore, the high correlation may perform better than lower
correlation, since the lower correlation would waste partial gain
when the number of nonzero eigenvalues is larger than $M$.

For MISO-MC-CDMA systems with precoded STFBC, there exist space,
time, frequency and multiuser resources available to promise system
reliable. Remarkably, adaptive precoding and optimal power
allocation can provide a better performance over traditional fixed
transmission strategies under the same condition, especially at
lower SINR, which will be confirmed later by the simulation results.

\section{The Special Transmission Scenarios}
Although the closed-form solutions of the optimization problem given
by (\ref{equ18}) have been obtained, the power allocation matrix of
the optimal precoder should be solved through iterative
water-filling, which arises high computational complexity at BS,
especially with numerous MUs. For some special scenarios, we can
adopt simple algorithms to achieve optimal or near optimal
performance, which promise our proposed transmission schemes more
feasible. In what follows, we discuss these special scenarios in
detail, respectively.

\subsection{Numerous Delay Taps, Long Spread Code and
high SINR} It is shown in (\ref{equ18}) that the number of delay
taps $L$ and the length of spread code $N_v$ have a positive impact
on the upper bound of the average PEP. As defined above, $L$ is
represented as the number of resoluble transmission paths. The
number of resoluble path deeply depends on the propagation
environment, for example, typical urban channel model (TUx) for UMTS
specified by 3GPP has 20 delay taps and channel no. 6 for DRM system
produced by ETSI has 5 resoluble paths. Similarly, the length of
spread code $N_v$ may be large according to the system requirement.
In addition, the transmit SINR $\eta$ can be also very high due to
large transmit power. We define $\Gamma=\eta LN_v$ as the equivalent
transmit SINR. With the $L$, $N_v$ and $\eta$, $\Gamma$ becomes
quite large. In this context, we have the following theorem:

\emph{Theorem 3}: In multiuser MISO-MC-CDMA systems, at high
effective SINR region due to large $L$, $N_v$ or $\eta$, for the
optimal procoder, $\textbf{V}_0=\textbf{V}_{\textbf{R},0}$ is used
to adapt to the channel condition,
$\textbf{U}_{0}=\textbf{V}_{\textbf{A},0}$ makes the precoder
matched with space time codeword to be transmitted, and
$\textbf{D}_0=\sqrt{1/M}\textbf{I}_M$, namely equal power
allocation, can achieve near optimal performance.

\begin{proof} The objective function in (\ref{equ18}) can be written
as
$\det\left(\textbf{I}_{N_t}+\Gamma\textbf{F}_{0}\textbf{A}_0\textbf{F}_{0}^{H}\textbf{R}_{0}\right)$.
Note that, when the effective SINR $\Gamma\rightarrow\infty$, then
the term $\textbf{I}_{N_t}$ in the objective function can be omitted
with respect to the dominating term
$\Gamma\textbf{F}_{0}\textbf{A}_0\textbf{F}_{0}^{H}\textbf{R}_{0}$.
Thus in the limit, the optimization problem is equivalent to
maximizing
$\det\left(\textbf{I}_{N_t}+\Gamma\textbf{F}_{0}\textbf{A}_0\textbf{F}_{0}^{H}\textbf{R}_{0}\right)$,
subject to the power constraint, which can be written as
(\ref{equ35}) at the top of next page.
\begin{figure*}
\begin{equation}
(\textbf{V}_0, \textbf{D}_0,
\textbf{U}_0)=\arg\max\limits_{\textbf{V}_0\in\textbf{U}^{N_t\times
N_t},\textbf{D}_0\in\textbf{D}^{M\times M},
\textbf{U}_0\in\textbf{U}^{M\times
M}}\det\left(\Gamma\textbf{V}_{0}\textbf{D}_{0}\textbf{U}_{0}^{H}\textbf{V}_{\textbf{A},0}
\textbf{D}_{\textbf{A},0}\textbf{V}_{\textbf{A},0}^H\textbf{U}_{0}
\textbf{D}_{0}\textbf{V}_{0}^{H}\textbf{V}_{\textbf{R},0}\textbf{D}_{\textbf{R},0}\textbf{V}_{\textbf{R},0}^H\right).\label{equ35}
\end{equation}
\end{figure*}
According to Hadamard's inequality, in order to maximize the above
objective function, the following conditions must be satisfied:
$\textbf{V}_0=\textbf{V}_{\textbf{R},0}$ and
$\textbf{U}_{0}=\textbf{V}_{\textbf{A},0}$. So the objective
function reduces to
\begin{equation}
\textbf{D}_0=\arg\max\limits_{\textbf{D}_0\in\textbf{D}^{M\times
M}}\det\left(\Gamma\textbf{D}_{0}^{2}\textbf{D}_{\textbf{A},0}\textbf{D}_{\textbf{R},0}^{'}\right),\nonumber
\end{equation}
whereafter, let
$\textbf{D}_{0}=\textmd{diag}\{d_1,\cdots,d_i,\cdots,d_M\}$, the
power allocation problem is equivalent to the following optimization
problem:
\begin{eqnarray}
J_2:
\max\limits_{\textbf{D}_0}&&{\det\left(\Gamma\textbf{D}_{0}^{2}\textbf{D}_{\textbf{A},0}\textbf{D}_{\textbf{R},0}^{'}\right)}={\det\left(\Gamma\textbf{D}_{\textbf{A},0}\textbf{D}_{\textbf{R},0}^{'}\right)\prod\limits_{i=1}^{M}d_i^2}\nonumber\\
\textmd{s.t.}&& \sum\limits_{i=1}^{M}d_i^2=1\nonumber\\
&&d_i\geq0,\quad i=1,\cdots,M.\nonumber
\end{eqnarray}
This optimization problem $J_2$ can be solved based on the fact that
$\prod\limits_{i=1}^{M}d_i^2\leq\left(\sum\limits_{i=1}^M\frac{d_i^2}{M}\right)^M=M^{-M}$,
and the equality holds if and only if $d_1^2=\cdots d_M^2=1/M$, so
we have $\textbf{D}_{0}=\sqrt{1/M}\textbf{I}_M$.
\end{proof}

\subsection{Low SINR}
By using Taylor expansion, $\det\left(\textbf{I}_{N_t}+\eta
LN_v\textbf{F}_{0}\textbf{A}_0\textbf{F}_{0}^{H}\textbf{R}_{0}\right)$
can be expressed as $1+\eta
LN_v\textmd{tr}\left(\textbf{D}_{0}^{2}\textbf{D}_{\textbf{A},0}\textbf{D}_{\textbf{R},0}^{'}\right)+\mathcal{O}(\eta)$,
where $\mathcal{O}(\eta)$ denotes the high order term of the SINR
$\eta$. When $\eta$ is sufficiently small, the high order term
$\mathcal{O}(\eta)$ becomes negligible. Then, the objective function
is reduced to $1+\eta
LN_v\textmd{tr}\left(\textbf{D}_{0}^{2}\textbf{D}_{\textbf{A},0}\textbf{D}_{\textbf{R},0}^{'}\right)$.
In this context, the power allocation can be described as the
following optimization problem:
\begin{eqnarray}
J_3:\max\limits_{\textbf{D}_0}&&\textmd{tr}\left(\textbf{D}_{0}^{2}\textbf{D}_{\textbf{A},0}\textbf{D}_{\textbf{R},0}^{'}\right)\nonumber\\
\textmd{s.t.}&&\textmd{tr}(\textbf{D}_{0}^{2})=1\nonumber\\
&&\textbf{D}_{0}\succeq0.\nonumber
\end{eqnarray}
By solving this problem, we could get the following theorem:

\emph{Theorem 4}: In multiuser MISO-MC-CDMA systems, at low SINR
region, for the optimal procoder,
$\textbf{V}_0=\textbf{V}_{\textbf{R},0}$ is used to adapt space time
codeword to the channel condition,
$\textbf{U}_{0}=\textbf{V}_{\textbf{A},0}$ makes the precoder
matched with space time codeword to be transmitted, and
$\textbf{D}_0=\textmd{diag}\{0,\cdots,1,\cdots,0\}$, where the
position of $1$, represented as $m^{\star}$, satisfies the below
condition: $m^{\star}=\arg\max\limits_{1\leq m\leq
N_t}\left[\textbf{D}_{\textbf{A},0}\textbf{D}_{\textbf{R},0}^{'}\right]^{(m)}$,
$[\textbf{B}]^{(m)}$ represents the $m$th diagonal element of
$\textbf{B}$. Generally speaking, distributing all transmit power to
the $m^{\star}$th spatial subchannel achieves the optimal
performance asymptotically.

\begin{proof} Let
$\textbf{D}_{\textbf{A},0}=\textmd{diag}\{d_{\textbf{A},1},\cdots,d_{\textbf{A},m},\cdots,d_{\textbf{A},M}\}$
and
$\textbf{D}_{\textbf{R},0}^{'}=\textmd{diag}\{d_{\textbf{R},1},\cdots,d_{\textbf{R},m},\cdots,d_{\textbf{R},M}\}$.
The optimization problem $J_3$ can be rewritten as
\begin{eqnarray}
J_4:\max\limits_{\textbf{D}_0}&&\sum\limits_{m=1}^{M}d_m^2d_{\textbf{A},m}d_{\textbf{R},m}\nonumber\\
\textmd{s.t.}&&\sum\limits_{m=1}^{M}d_m^2=1\nonumber\\
&&d_m\geq0,\quad m=1,\cdots, M.\nonumber
\end{eqnarray}
Clearly, the optimization problem $J_4$ is a linear programming
problem. Considering the feasible set is a polyhedron, the basic
feasible solution must be one of the vertices of the polyhedron. For
this problem, there are $M$ poles in the form of
$(0,\cdots,1,\cdots,0)$ comprising one 1 and $M-1$ zeros. Obviously,
only when $d_{m^{\star}}=1,
d_1=\cdots=d_{{m}^{\star}-1}=d_{{m}^{\star}+1}=\cdots=d_{N_t}=0$,
where $m^{\star}=\arg\max\limits_{1\leq m\leq
N_t}d_{\textbf{A},m}d_{\textbf{R},m}$, the objective function is
maximal.
\end{proof}

\section{Simulation Results}
In order to examine the performance of the proposed system
framework, we present some representative simulation results in
various scenarios. The main simulation parameters, referred to the
IEEE 802.16e system simulation requirements \cite{802.16m}, are
configured as seen in Tab.1. Moreover, we adopt the ITU channel
model suggested by the draft. Data is modulated by 16-QAM and
transmitted frame by frame. In the duration of a data frame, the
channel is assumed to keep constant and fades in inter-frames
according to the famous Jakes model. Note that we set the maximal
doppler spread as 50Hz, then the channel variety is not so fast and
the statistical information can be estimated perfectly. In what
follows, we construct the transmit correlation matrix as follows:
$\textbf{R}_0(m,n)=\kappa^{|m-n|}$, where $\textbf{R}_0(m,n)$ is the
$(m,n)$th element of correlation matrix $\textbf{R}_0$ and $\kappa$
is the so called correlation coefficient. For convenience of
implementation, Alamouti code is used in all simulation cases.

\newcommand{\tabincell}[2]{\begin{tabular}{@{}#1@{}}#2\end{tabular}}
\begin{table}\centering
\caption{System Parameter Table} \label{Tab1}
\begin{tabular*}{7.8cm}{|c|c|c|}\hline
Parameter & Description & Value\\
\hline $f_c$& Carrier Frequency & 2.5 GHz\\
\hline BW & Total Bandwidth & 10 MHz\\
\hline $N_c$ & Number of Sub-Carriers & 1024\\
\hline $F_s$ & Sampling Frequency & 11.2 MHz\\
\hline $\triangle{f}$ & Sub-Carrier Space & 10.9375 KHz\\
\hline CP & Cycle Prefix Length & 128\\
\hline $T_s$ & OFDM Symbol Length & 102.86 us\\
\hline $T_F$ & Frame Length & 5 ms\\
\hline $N_v$ & Orthogonal Spread Factor & 8\\
\hline $N_t$ & Number of Transmit Antennas & 4\\
\hline $N_u$ & Number of MUs on Each Carrier & 4\\
\hline
\end{tabular*}
\end{table}

\subsection{Different Transmission Schemes}
\begin{figure}[h] \centering
\includegraphics [width=0.5\textwidth] {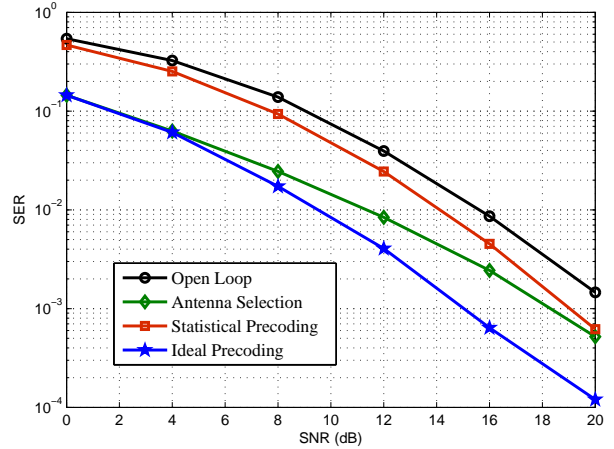}
\caption {SER performance comparison with different transmission
schemes.} \label{Fig2}
\end{figure}

At first, we compare the SER performance of four different
transmission schemes, namely open loop, antenna selection,
statistical precoding and ideal precoding with $\kappa=0.3$. As the
names imply, open loop means that the transmitter has no channel
statistical information and the Alamouti code are directly
transmitted without precoding, antenna selection chooses two optimal
antennas from all transmit antenna over all carriers for the
transmission of Alamouti code based on the feedback information from
the MUs, statistical precoding is the proposed precoding scheme, and
ideal precoding denotes that the precoding direction matrix and
power allocation matrix over each sub-carrier are designed based on
full channel instantaneous information. As seen in Fig.\ref{Fig2},
in comparison with open loop, statistical precoding has an obvious
performance gain. For example, at the SER of $10^{-2}$, statistical
precoding provides more than 1.5dB gains. Moreover, with the
increase of SNR, the performance gain increases accordingly.
Although antenna selection performs better than statistical
precoding significantly, it is difficult to apply in practical
multi-carrier systems, because it needs to update selection
information within a channel correlation time over all subcarriers,
which is quite unbearable. Fortunately, statistical precoding can
asymptotically approach the performance of antenna selection in high
SNR region. From the perspectives of feedback amount and
computational complexity, open loop does not require any feedback
information and has the lowest complexity. Antenna selection needs
to convey the selection information during each correlation time and
has a lower complexity. With respect to antenna selection, the
proposed statistical precoding has a relatively higher complexity,
but requires less feedback amount, since channel statistical
information varies very slowly. Additionally, it does not need to
convey the information over all sub-carriers as mentioned earlier.
Ideal precoding has the highest complexity and requires infinite
feedback amount. Thereby, statistical precoding is a preferred
choice in multi-carrier systems.

\subsection{Different Power Allocation}
\begin{figure}[h] \centering
\includegraphics [width=0.5\textwidth] {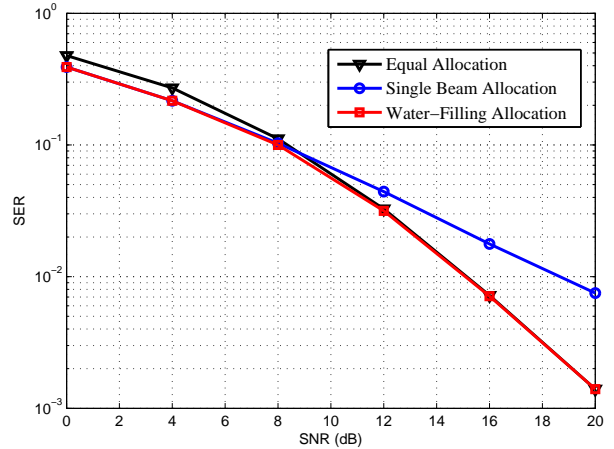}
\caption {SER performance comparison with different power allocation
methods.} \label{Fig3}
\end{figure}

In this subsection, we investigate the effect of power allocation
methods on the SER performance with $\kappa=0.1$. As seen in
Fig.\ref{Fig3}, at low SNR, under 8dB, single beam allocation nearly
performs the same as water-filling allocation, but there is a large
gap at high SNR. For example, water-filling allocation has more than
3dB gain at the SER of $10^{-2}$. On the other hand, with the
increase of SNR, the performance of equal allocation is
asymptotically close to that of water-filling allocation, which is
consistent with our theoretical claim. Therefore, we can replace
water-filling allocation with single beam allocation in low-SNR
regime and equal allocation in high-SNR regime, so that computation
complexity can be reduced significantly, while there is no obvious
SER performance loss.

\subsection{Different Correlation Coefficients}
\begin{figure}[h] \centering
\includegraphics [width=0.5\textwidth] {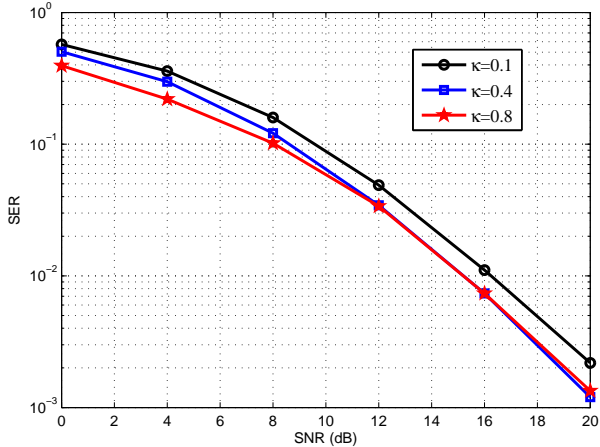}
\caption {SER performance comparison with different correlation
coefficients.} \label{Fig4}
\end{figure}

Fig.\ref{Fig4} illustrates the SER performance with different
correlation coefficients. For $M<N_t$, the performance with
$\kappa=0.1$ is clearly worse than that with $\kappa=0.4$. For
example, there is about 1dB performance gap at the SER of $10^{-2}$.
It is because partial channel gain is wasted under low correlation
condition, which confirms our theoretical claim earlier. With the
increase of $\kappa$, there is still a performance gain in low-SNR
regime, but not in high-SNR regime, because the coding gain from
$\Pi_{m=1}^{M}d_{\textbf{R},m}^{'}$ can be neglected when SNR is
large enough.

\section{Conclusions}
In this paper, we proposed a multiuser MISO-MC-CDMA system framework
for broadband wireless communication systems. This configuration
utilizes the space-time-frequency signal to obtain multi-diversity
gains, so that it can provide the higher transmission reliability.
To further improve the system performance and reduce receive
complexity, we exploit the channel information to linear precode the
space-time codeword on each sub-carrier. Under the consideration of
the amount of feedback, we propose the design method of the optimal
precoder, including precoding direction matrix and power allocation
matrix, in the sense of statistical average. For some special
transmission scenarios, we have given the corresponding simple
design method to reduce computational complexity at BS. Numerous
simulation results confirm our theoretical claims.

\end{document}